\documentclass[%
 reprint,
 notitlepage,
superscriptaddress,
nofootinbib,
 amsmath,amssymb,
 aps,
twocolumn, 
]{revtex4-2}

\usepackage{tikz}
\usepackage{graphicx}
\usepackage{dcolumn}
\usepackage{bm}

\usepackage{xcolor}
\usepackage{tabularx}
\usepackage[colorlinks]{hyperref}
\hypersetup{
    colorlinks,
    linkcolor={red!50!black},
    citecolor={blue!50!black},
    urlcolor={blue!80!black}
}
\usepackage[normalem]{ulem}
\usepackage{physics}
\usepackage{comment}
\usepackage{multirow}
\usepackage{subfigure}

\newcommand{\ie}{{\it i.e.}}

\newcolumntype{Y}{>{\centering\arraybackslash}X}
\newcolumntype{Z}[1]{>{\centering\arraybackslash}m{#1}}
\usepackage{cellspace}
\begin{document}

\title{Constraints on Heavy Asymmetric and Symmetric Dark Matter  \\
from the Glashow Resonance}

\author{Qinrui Liu}
\email{qinrui.liu@queensu.ca}
\affiliation{
Department of Physics, Engineering Physics and Astronomy,\\
Queen’s University, Kingston ON K7L 3N6, Canada
}
\affiliation{%
Arthur B. McDonald Canadian Astroparticle Physics Research Institute, Kingston ON K7L 3N6, Canada
}
\affiliation{Perimeter Institute for Theoretical Physics, Waterloo ON N2L 2Y5, Canada}

\author{Ningqiang Song}
\email{songnq@itp.ac.cn}
\affiliation{Institute of Theoretical Physics, Chinese Academy of Sciences, Beijing, 100190, China}

\author{Aaron C. Vincent}
\email{aaron.vincent@queensu.ca}
\affiliation{
Department of Physics, Engineering Physics and Astronomy,\\
Queen’s University, Kingston ON K7L 3N6, Canada
}
\affiliation{%
Arthur B. McDonald Canadian Astroparticle Physics Research Institute, Kingston ON K7L 3N6, Canada
}
\affiliation{Perimeter Institute for Theoretical Physics, Waterloo ON N2L 2Y5, Canada}


\begin{abstract}

The decay of asymmetric dark matter (ADM) can lead to distinct neutrino signatures characterized by an asymmetry between neutrinos and antineutrinos. In the high-energy regime, the Glashow resonant interaction $\bar{\nu}_{e} + e^{-} \rightarrow W^{-}$ yields an increase in sensitivity to the neutrino flux, and stands out as the only way of discerning the antineutrino component in the diffuse high-energy astrophysical neutrino flux. This offers a unique opportunity in the search for dark matter with masses above the PeV scale. We examine the neutrino signal stemming from ADM decay and set the first stringent constraints on ADM lifetime $\tau_X$. For ADM with mass $m_X\gtrsim 10$~PeV, we {exclude} $\tau_X\lesssim 10^{29}$s using the data from the recent IceCube search for Glashow resonance events.  Our projections further show that sensitivities at the forthcoming IceCube-Gen2 could approach $10^{30}$s, depending on the decay channel.  The current constraints on symmetric dark matter decay to neutrinos are also improved by up to a factor of 3 thanks to the Glashow resonance.

\end{abstract}

\maketitle

\textbf{\textit{Introduction}} --- High-energy astrophysical neutrinos in the TeV-PeV range have been detected in large Cherenkov telescopes, most notably IceCube~\cite{IceCube:2013low}. Detection is primarily through deep inelastic scattering (DIS) with nucleons, producing event morphologies that are recognized as showers (or cascades), tracks, or double cascades in or near the detector. Although the morphologies allow some reconstruction of the flavor of high-energy neutrinos, the identification of the neutrino charge conjugation-parity ($CP$) eigenstates in these interactions remains impossible. The only way to pin down the antineutrino component from the total neutrino flux at high energies is the so-called Glashow resonance (GR)~\cite{Glashow:1960zz}, i.e., the resonant scattering of $\bar{\nu}_e$ on electrons in the detector $\bar{\nu}_{e} + e^{-} \rightarrow W^{-}$, which dramatically enhances the cross section for neutrino energies near {$E_{\rm GR}=6.3$}~PeV. 

Such an event was first reported by IceCube as a partially contained cascade with a reconstructed energy of 6.05$\pm$0.72~PeV~\cite{IceCube:2021rpz}. The detection of GR events with neutrino telescopes such as IceCube can grant a deeper understanding of astrophysical processes~\cite{Brown:1981ns,Anchordoqui:2004eb,Bhattacharya:2011qu,Xing:2011zm,Barger:2012mz, Bhattacharya:2012fh,Anchordoqui:2014yva,Barger:2014iua,Palladino:2015uoa,Anchordoqui:2016ewn,Biehl:2016psj,Huang:2019hgs,Huang:2023yqz,Liu:2023lxz,Huang:2023mgt}, and opens the door to new physics searches, e.g. neutrino decay and Lorentz invariance violation~\cite{Stecker:2014oxa,Shoemaker:2015qul,Bustamante:2020niz,Xu:2022svm}. In a variety of models, dark matter (DM) interacts primarily with neutrinos, enabling the decay or annihilation of DM into neutrinos~\cite{Blennow:2019fhy,Alvey:2019jzx,Baumholzer:2019twf,Boehm:2006mi,Escudero:2016tzx,Farzan:2012sa,Farzan:2014gza,Patel:2019zky,Garcia-Cely:2017oco,Coy:2020wxp}. In such scenarios, the direct detection of DM becomes challenging, although indirect searches for neutrino-DM elastic scattering benefit from an enhanced cross section at high energies~\cite{Arguelles:2017atb}. Searches for neutrino signals from DM annihilation  \cite{Arguelles:2019ouk} or decay \cite{,Arguelles:2022nbl} have also led to strong constraints on these models. Decay to neutrinos is particularly well-constrained at very high energies thanks to the growing neutrino-nucleus cross section and the presence of gamma rays from electroweak corrections. 

\begin{figure}[tp!]
    \centering
    \includegraphics[width=\columnwidth]{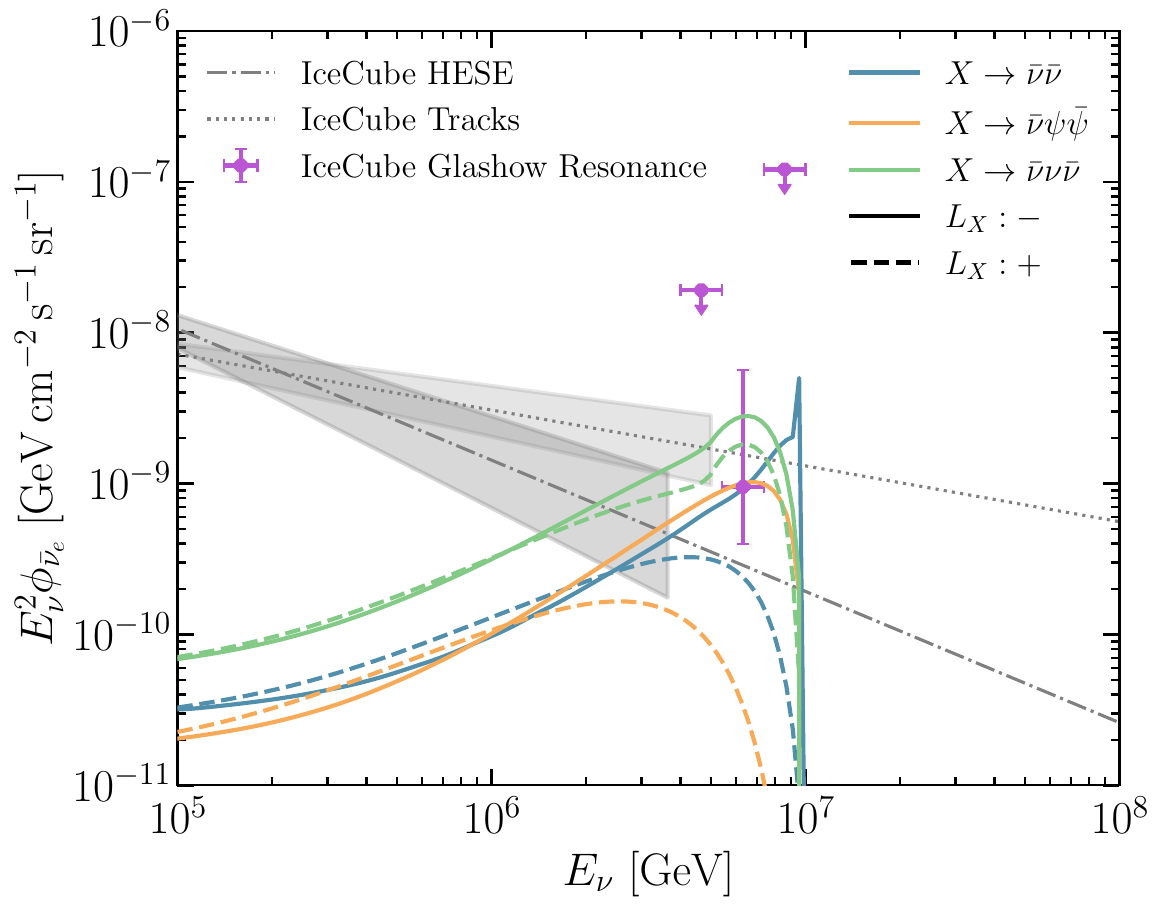}
    \caption{Expected {$\bar{\nu}_e$} spectra of with the three decay modes, \ie~$X\rightarrow \bar{\nu}\bar{\nu}(\nu\nu)$ (blue), $X\rightarrow \bar{\nu}(\nu)+\psi\bar{\psi}$ (orange) and $X\rightarrow \bar{\nu}(\nu)+\nu\bar{\nu}$ (green) for $m_X=20~\rm{PeV}$ when the lifetime $\tau_X=10^{29}\rm{s}$. A solid line corresponds to the mode with a negative lepton number and the dashed line corresponds to a positive lepton number. The IceCube GR observation~\cite{IceCube:2021rpz} is shown as purple points. Gray lines are IceCube diffuse astrophysical neutrino measurements of high-energy starting events (dash-dotted)~\cite{IceCube:2020wum} and through-going muon tracks (dotted)~\cite{IceCube:2021uhz} assuming equal contribution of each flavor. For IceCube diffuse astrophysical neutrino measurements shown here, we assume $\bar{\nu}_e$ constitutes 1/6 of the total flux.}
    \label{fig:flux_all}
\end{figure}

As the GR breaks the degeneracy between neutrinos and antineutrinos, it is a sensitive probe of asymmetric dark matter (ADM), an alternative to the widely studied weakly interacting massive particles (WIMPs). Asymmetric models were proposed to connect the relic abundance of DM to the baryon asymmetry in the Universe~\cite{Kaplan:2009ag}, but are not generically tied to baryogenesis. If the dark sector's asymmetry is linked to the standard model (SM) sector, then decay could give rise to an asymmetry between SM decay products \cite{Nardi:2008ix,Chang:2011xn, Masina:2011hu,Masina:2012hg,Masina:2013yea,Feng:2013vva,Zhao:2014nsa}, including neutrinos~\cite{Feldstein:2010xe,Fukuda:2014xqa}. Figure~\ref{fig:flux_all} shows predictions from a few such scenarios overlaid on observational data. These scenarios will be described later in this text.

In this work, we will exploit the enhanced cross section in the Glashow region to 1) improve on existing constraints on the decay of \textit{symmetric} dark matter to $\bar \nu \nu$, and 2) derive new constraints on \textit{asymmetric} dark matter decay to a number of neutrinophilic channels, setting stringent constraints on the ADM lifetime at $\tau_X\lesssim 10^{29}$~s for ADM mass $m_X\gtrsim 10$~PeV. Neutrinos produced from electroweak radiation preserve information about the asymmetry, and allow us to cover a mass range between $10^{7}$~GeV and $10^{10}$~GeV. We will present constraints from currently available IceCube data, and projected sensitivities from the upcoming IceCube-Gen2. 


\textbf{\textit{Heavy Asymmetric Dark Matter Decay}} --- Although some of the first ADM models predicted new particles in the 5-15 GeV mass range \cite{Kaplan:2009ag}, models exist spanning a vast range of scales, and heavy ADM in particular has been widely explored~\cite{Nardi:2008ix,Chang:2011xn,Masina:2011hu,Thomas:1995ze,Kitano:2004sv,Unwin:2012rp,Zhao:2014nsa,Asadi:2021yml}. Such constructions can be naturally realized in various scenarios, e.g., if the SM and the dark sector were never in thermal equilibrium and they both originate from the decay of a heavy mother particle with nonzero primordial baryon/lepton number asymmetry~\cite{Thomas:1995ze,Kitano:2004sv,Unwin:2012rp}, or the thermally produced ADM goes through self-annihilating process producing lighter dark particles and depleting its abundance to the level observed today~\cite{Zhao:2014nsa,Asadi:2021yml}. The mass of  {ADM} is effectively unconstrained in these scenarios, so we treat $m_X$ as a free parameter.

\renewcommand{\arraystretch}{1.0}
\begin{table*}[]

    \centering
    \begin{tabularx}{0.7\textwidth}{Z{1.5cm}Z{2.5cm}Z{2.5cm}Z{2.5cm}Z{2.5cm}}
    \hline
    
    \centering

      \rule{0pt}{15pt}Benchmark & B1 (scalar $X$) & B2 (scalar $X$) & B3 (fermion $X$) & B4 (fermion $X$)\\[5pt]

      \hline
      \rule{0pt}{15pt}$\mathcal{O}_{X\rightarrow \nu}$   & $\dfrac{1}{\Lambda}X\psi L\Phi$   & 
       $\dfrac{1}{\Lambda^2}X(L\Phi)^2$ &   $\dfrac{1}{\Lambda^2}XL\psi^2$ &  $\dfrac{1}{\Lambda^2}X LL\nu^c$ \\[5pt]

      \hline
      \centering
      \rule{0pt}{15pt} Decay    &$X \rightarrow \bar{\nu}\psi/\nu\psi$   & $X \rightarrow \bar{\nu}\bar{\nu}/\nu\nu$   & $X \rightarrow \bar{\nu}\psi\bar{\psi}/\nu\psi\bar{\psi}$  & $X \rightarrow \bar{\nu}\nu\bar{\nu}/\nu\nu\bar{\nu}$ \\[5pt]
    \hline
    \end{tabularx}

    \caption{Operators of the lowest dimensions and the corresponding neutrino-dominant decay modes of dark matter $X$. $L$ is the Standard Model lepton doublet, and $\psi$ is a dark fermion or SM particle.}
    \label{tab:decay}
\end{table*}

ADM typically carries $B-L$ (baryon minus lepton) numbers, and can transfer an asymmetry between the dark sector and SM via operators of the form
$
    \mathcal{O}_{\mathrm{ADM}} = \Lambda^{4-m-n}\mathcal{O}_X \mathcal{O}_\mathrm{SM},
$
where the dark sector $\mathcal{O}_X$ and SM $\mathcal{O}_\mathrm{SM}$ operators have dimensions $m$ and $n$ respectively with $\Lambda$ being the new physics scale~\cite{Feldstein:2010xe,Zhao:2014nsa}  We intend to be model-independent and focus on a set of representative benchmark operators that lead to neutrino decay, as listed in Table.~\ref{tab:decay}. We focus on dominant lowest-dimension operators that produce distinct GR signals depending on the $B-L$ charge of the DM. For all those scenarios, DM $X$ carries a lepton number $L_X$. In Table.~\ref{tab:decay}, $L$ is the SM lepton doublet and $\Phi$ is a heavy scalar doublet which can be integrated out~\cite{Feldstein:2010xe}. A scalar DM $X$ can decay through the two-body process to a neutrino and a dark fermion $\psi$, i.e., $X \rightarrow \bar{\nu}(\nu)+\psi$ (B1) or two neutrinos $X \rightarrow \bar{\nu}\bar{\nu} (\nu\nu)$ (B2). The last two operators
can arise from the operator $\bar{X}l\phi$ and $\frac{1}{\Lambda^{3n-3}}\phi \psi^n$ with the heavy scalar field $\phi$ integrated out~\cite{Hiroshima:2017hmy,Fukuda:2014xqa}. This leads to a multibody decay process for the fermionic DM $X\rightarrow \bar{\nu}(\nu)+2n\psi$. At the lowest order $n=1$, the decay produces a neutrino or antineutrino, depending on the lepton number carried by $X$ (B3). The fermion $\psi$ could be a new particle, or from the SM; if $\psi=\nu$  then $X$ decays to three neutrinos (B4).

\textbf{\textit{$\bar{\nu}_e$ Flux from ADM}} --- Assuming ADM makes up all DM, we compute the expected high-energy $\bar{\nu}_e$ flux from ADM in the Galactic halo and beyond the Milky Way via decay channels discussed above.

The averaged contribution from the Galactic halo with a solid angle coverage $\Omega$ for DM with mass $m_X$ and lifetime $\tau_X$ can be written as 
\begin{equation}\label{eq:galatic}
    \frac{d\Phi^{\mathrm{gal.}}_{\bar{\nu}_e}}{dE_\nu} = \frac{1}{4\pi m_{X}\tau_{X}}\sum_\alpha^3 \frac{dN^{\mathrm{ch}}_{\bar{\nu}_\alpha}}{dE_\nu}\frac{\mathcal{D}\left(\Omega\right)}{\Omega}P_{\bar{\nu}_\alpha\rightarrow \bar{\nu}_e},
\end{equation}
where $\mathcal{D}=\int d\Omega\int_{los}\rho_X\left[r(s,l,b)\right] ds$ represents the column density of DM along the line of sight (los) over the solid angle $\Omega=4\pi$. We adopt the generalized Navarro-Frenk-White (gNFW) profile for $\rho_X$ with a slope parameter 1.2, local DM density $0.4~\rm{GeV/cm^3}$ and a scale radius 20~kpc~\cite{deSalas:2019pee}. The galactocentric distance $r=\sqrt{s^2+R^2_{\odot}-2sR_\odot{\rm{cos}}\,l\,{\rm{cos}}\,b}$, where $l$ and $b$ correspond to the Galactic latitude and longitude. Switching from the gNFW profile to a cored profile only introduces a $\lesssim 20\%$ difference in $\mathcal{D}$. We assume a universal production of neutrino flavor $\alpha \in \{e,\mu,\tau\}$, and $dN^{\mathrm{ch}}_{\bar{\nu}_\alpha}/dE_\nu$ is the energy spectrum of neutrinos produced in the decay of a single DM particle via a specific decay channel to $\bar{\nu}_\alpha$. We also incorporate the neutrino oscillation probability after production, which leads to a $\sim 1/3$ contribution from each flavor to the total flux~\cite{Song:2020nfh}. In practice, we adopt the best-fit oscillation parameters from \texttt{NuFIT5.3}~\cite{Esteban:2020cvm,nufit5.3} to compute this probability. 

The extragalactic DM contribution to the isotropic neutrino flux is
\begin{equation}\label{eq:extragalatic}
    \frac{d\Phi^{\mathrm{ext.\,gal.}}_{\bar{\nu}_e}}{dE_\nu} = \frac{\Omega_{X}\rho_c}{4\pi m_{X}\tau_{X}}\sum^3_\alpha \int_0^{\infty}\frac{dN^{ch}_{\bar{\nu}_\alpha}}{dE'_\nu}\frac{dz}{H(z)}P_{\bar{\nu}_\alpha\rightarrow \bar{\nu}_e},
\end{equation}
where $\Omega_X = 0.229$ is the DM density in the Universe, $\rho_c$ is the Universe critical density and $z$ is the redshift. $E'_\nu\equiv(1+z)E_\nu$ is the redshifted neutrino energy and $H(z)=H_0\sqrt{\left(1+z\right)^3\Omega_m + \Omega_\Lambda}$ is the Hubble expansion rate, for which we use $H_0=67.4\,\rm{km\,s^{-1}\,Mpc^{-1}}$, $\Omega_m=0.315$, and $\Omega_\Lambda=0.685$~\cite{Planck:2018vyg}.

We sum the galactic contribution in Eq.~\eqref{eq:galatic} and the extragalactic contribution in Eq.~\eqref{eq:extragalatic} to find the angular-averaged neutrino flux from ADM decay. As the DM mass considered in this work is well above the electroweak scale, final-state radiation will modify the neutrino spectrum, softening the primary neutrino spectrum, and producing an additional lower energy $\nu$ and $\bar \nu$ component from electroweak showers. With these corrections, the energy spectrum of $\bar{\nu}_\alpha$ is
\begin{align}
\frac{dN^{\mathrm{ch}}_{\bar{\nu}_\alpha}}{dE_\nu}(E_\nu) = \sum_i \int_{E_\nu/m_X}^{1}\frac{1}{ym_X}\frac{df_i}{dy}D^{\bar{\nu}_\alpha}_i(x;ym_X)dy,
\end{align}
where the sum on $i$ goes over each (anti)neutrino produced in the uncorrected decay, including a sum over flavors. $f_i$ is the distribution function for decay product $i$, which we provide in the Supplementary Material. We implement the fragmentation functions $D^{j}_{i}(x;ym_X)$ from \texttt{HDMSpectrum}~\cite{Bauer:2020jay} where $y$ represents the fraction of energy taken away by particle $i$ in the DM mass  and $x\equiv E_j/(ym_X)$ is the energy fraction deposited to the final particle $j$ from the initial particle energy. The subsequent evolution such as hadronization and showering handled by \texttt{PYTHIA8.2}~\cite{Sjostrand:2006za,Sjostrand:2007gs,sjostrand:2014zea} is also included in the fragmentation functions. 

The $\bar{\nu}_e$ spectra from the benchmark decay modes are shown in Fig.~\ref{fig:flux_all}. The electroweak corrections tend to broaden the Dirac delta spectrum in two-body decay, and generate a $\bar{\nu}_e$ flux even when antineutrinos are absent in the initial decay products. If $X$ carries negative lepton number $L_X<0$ which tends to decay more to antineutrinos, the flux is more peaked at high energies. On the contrary, if $L_X>0$ the $\bar{\nu}_e$ spectrum is more suppressed at the highest energies and peaks at lower energies. At $E_{\bar{\nu}_e}/m_X\lesssim 10^{-2}$, the $\bar{\nu}_e$ spectral for $L_X>0$ and $L_X<0$ converge. This means that GR loses the ability to pin down the asymmetry and lepton number of $X$ when $m_X\gtrsim$~EeV.

\textbf{\textit{Current IceCube Data}} --- We set constraints on ADM with publicly available IceCube data~\cite{IceCube:2021rpz}. The measured flux is under the assumption of $(\nu:\bar{\nu})=(1:1)$ and a flavor ratio of ($\nu_e:\nu_\mu:\nu_\tau)= (1:1:1)$ at Earth. The segmented differential flux reported in Ref.~\cite{IceCube:2021rpz} covers the energy range [4, 10]~PeV assuming an $E^{-2}$ power-law spectrum within each bin. We integrate the reported differential flux limits over each bin to obtain $\Phi_{\mathrm{obs},i}$. The uncertainty can be similarly integrated to obtain $\sigma^2_{\Phi_{\mathrm{obs}},i}$.

The measured $\bar{\nu}_e$ flux receives contributions mainly from two sources: the diffuse astrophysical neutrino flux and the flux from DM decay. We ignore the atmospheric neutrino flux which is negligible next to the astrophysical flux at PeV energies. The main background is therefore from the isotropic astrophysical neutrino flux which is typically modeled as a power-law     ${d\Phi^{\mathrm{astro.}}_\nu}/{dE_\nu} = \phi_0 \left({E_\nu}/{100\mathrm{\, TeV}}\right)^{-\gamma}, $ where $\phi_0$ is the flux normalization and $\gamma$ is the spectral index. In this work, we fix the values to the best fits of the 7.5yr high-energy starting events (HESE) analysis~\cite{IceCube:2020wum}. We assume that $\bar{\nu}_e$ makes up 1/6 of the total astrophysical flux at Earth, corresponding to the equal-flavor  assumption used in the HESE analysis. This is consistent with a general scenario in which these neutrinos are produced from $\pi^\pm$ decays at the source and experience oscillation during the propagation. Since the statistics of PeV neutrinos are low, the spectral index plays an important role in the predicted event rate. We will discuss the effects of the spectral index and the $\bar{\nu}_e$ fraction when presenting the results.

To evaluate the limits on the DM lifetime, we define a Gaussian likelihood
\begin{align}
    -2\mathrm{ln}\mathcal{L}\left(\tau_X, m_X\right) = \sum_i^3\frac{\left(\Phi_{\mathrm{exp},i}-\Phi_{\mathrm{obs},i}\right)^2}{\sigma^2_{\Phi_{\mathrm{obs}},i}}\,,
\label{eq:obs_llh}
\end{align}
where we integrate the DM and astrophysical differential neutrino fluxes to obtain the expected flux $\Phi_{\mathrm{exp},i}$ within each bin. For each of the two bins where there are only upper limits, we take the measured value as 0 with the 68\% upper limit being the standard deviation and treat it as a half Gaussian distribution. 
The test statistic 
$
\mathrm{TS} =  2\mathrm{ln}[\mathcal{L}\left(\hat{\tau}_X, m_X\right)/\mathcal{L}\left(\tau_X, m_X\right)],    
$
where $\hat{\tau}_X$ is the value that maximizes Eq.~\eqref{eq:obs_llh}. Assuming Wilks's theorem holds, we obtain  the 90\% limit on the lifetime for each individual mass by finding the value of $\tau_X$ for which   $\mathrm{TS}=2.71$.

 \begin{figure}[htp!]
    \centering
    \subfigure
    {\includegraphics[width=0.95 \linewidth]{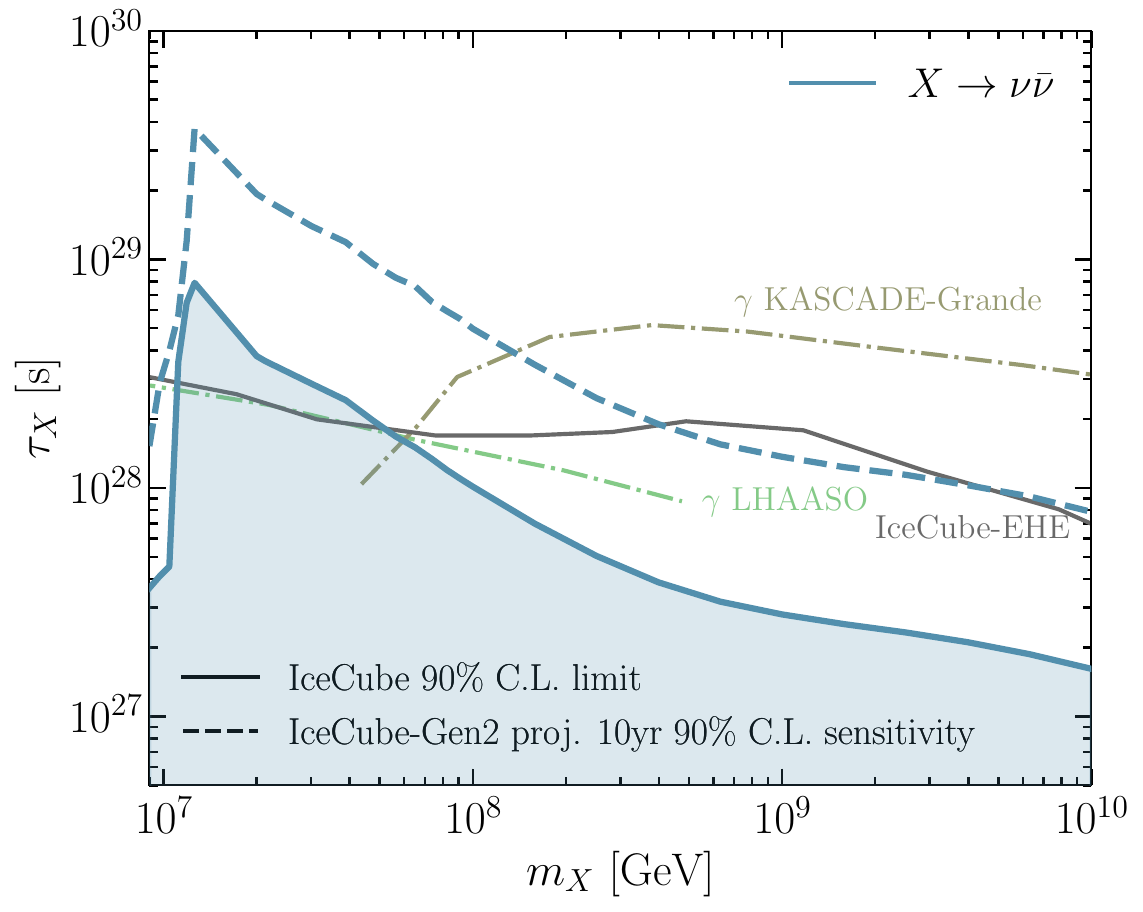}}\\ \vspace{-0.9cm}
    
    \subfigure
    {\includegraphics[width=0.95 \linewidth]{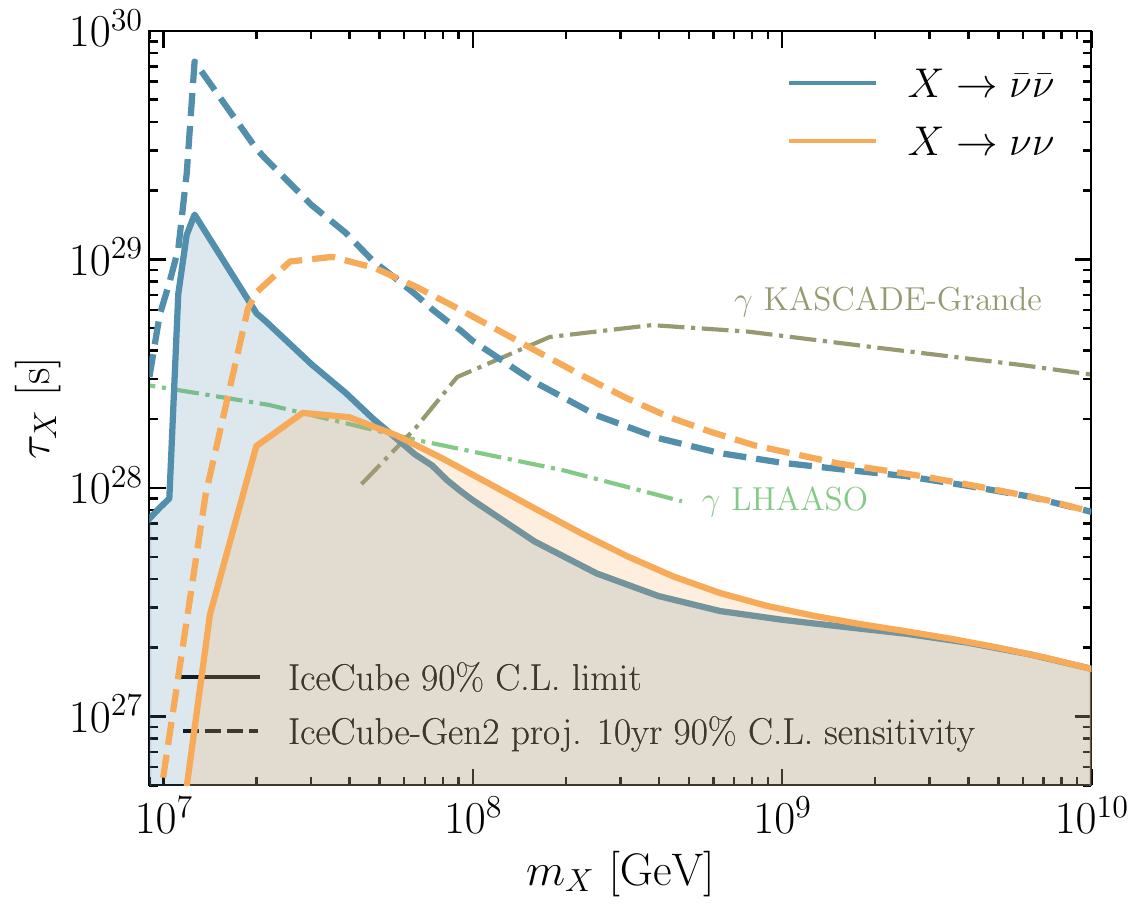}}\\
    \vspace{-0.9cm}
    \subfigure
    {\includegraphics[width=0.95 \linewidth]{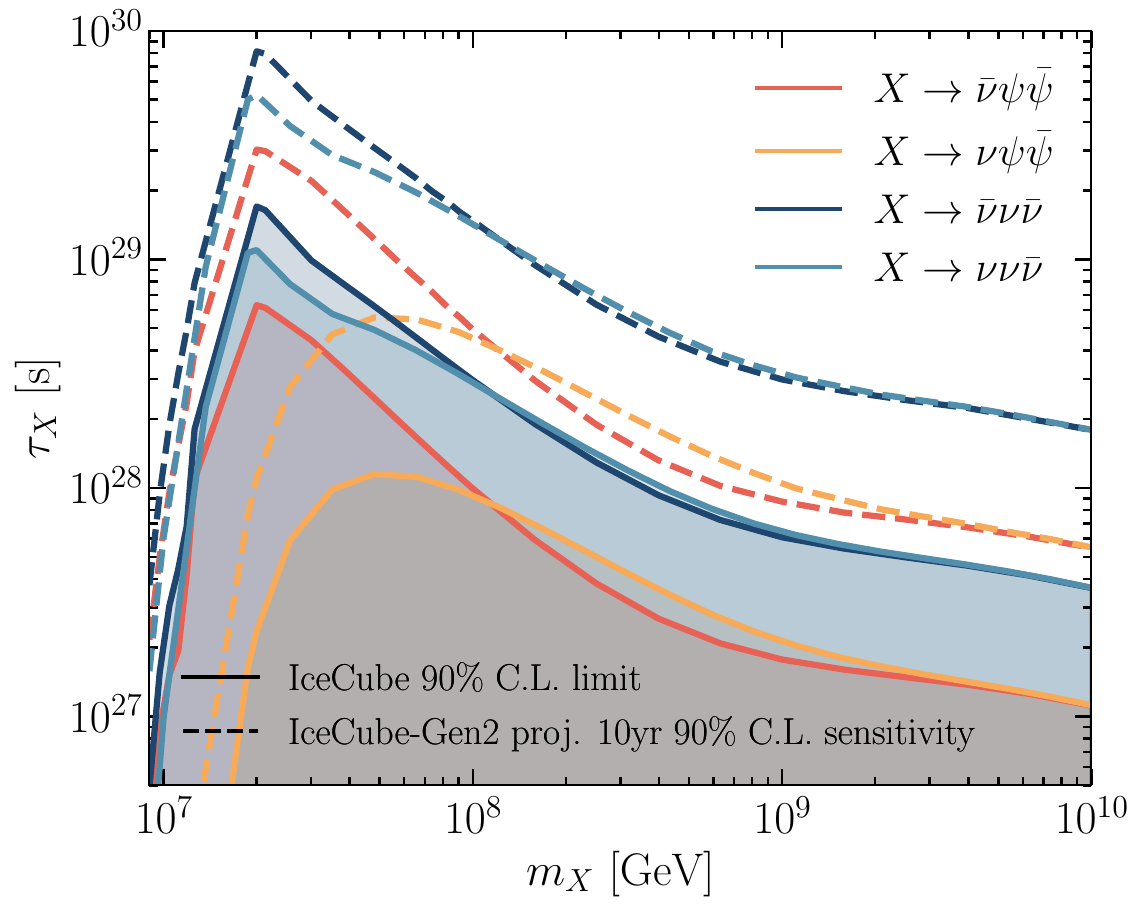}
    }    
    \caption{\textbf{\textit{Top:}}  90\% C.L. constraints on the lifetime of $X$ in $X\rightarrow \nu\bar{\nu}$ decay with the current IceCube observation of GR (shaded regions), and   the projected sensitivities in the future with 10~yr exposure of IceCube-Gen2 (dashed). We also show the corresponding limits with neutrino and gamma-ray observations from Ref.~\cite{Arguelles:2022nbl} derived from the diffuse gamma flux limit by KASCADE-Grande~\cite{KASCADEGrande:2017vwf} (light green) and the extremely-high-energy (EHE) astrophysical neutrino flux limit by IceCube~\cite{IceCube:2018fhm} (gray), as well as the LHAASO limit~\cite{LHAASO:2022yxw} (dark green). \textbf{\textit{Middle and bottom:}} 90\% C.L. constraints on the lifetime of ADM for the three decay modes. The gamma-ray limits here may also apply to the scenario of $X\rightarrow \bar{\nu}\bar{\nu} (\nu\nu)$. }
    \label{fig:lifetime_limits}
\end{figure}

\textbf{\textit{Future Projections}} --- We can also study the detectability of ADM decay at the next-generation neutrino telescopes. There is a suite of facilities under construction or proposed for the detection of high-energy astrophysical neutrinos with substantially larger exposure and event statistics. Here we focus on IceCube-Gen2, which will extend the current IceCube detector to a volume of 8~$\mathrm{km^3}$~\cite{IceCube-Gen2:2020qha}. We will discuss other next-generation experiments in the Supplementary Material.

We focus on the partially contained event selection \cite{IceCube:2021rpz}, which is the selection used for the IceCube GR observation and enlarges the effective area by approximately a factor of 2 in the GR energy window compared to the contained event selection (HESE, \cite{IceCube:2020wum}).  We build our projection on the expectation that GR events can be identified on an event-wise basis. As discussed in Ref.~\cite{Liu:2023lxz}, the dominant hadronic decay channel $W^-\rightarrow$ hadrons and the $W^-\rightarrow \mu^- + \bar{\nu}_\mu$ channel should be differentiable from DIS events from the energy deposition and morphology. 

We therefore perform a binned likelihood ratio analysis based on the Poisson likelihood of Glashow-induced cascades and tracks in each of the three bins defined in Ref.~\cite{IceCube:2021rpz}. To compute the sensitivity, the test statistic has TS = 2ln$\left[\mathcal{L}\left(\tau_X, m_X\right)/\mathcal{L}_\mathrm{null}\right]$ in which $\mathcal{L}_\mathrm{null}$ is the likelihood value of the null hypothesis where there is no contribution from the DM, \ie~$\tau_X$ is infinite. 

\textbf{\textit{Results}} --- 
We first present updated constraints on symmetric DM decay,  $X\rightarrow \nu\bar{\nu}$, in the top panel of Fig.~\ref{fig:lifetime_limits}. Thanks to the Glashow-enhanced cross section, our constraint on the dark matter lifetime surpasses the IceCube extremely-high-energy astrophysical neutrino flux limit~\cite{IceCube:2018fhm} in the $10-50$~PeV mass window, and is up to a factor of 3 stronger. 

{The constraints obtained with GR typically peak at $m_\chi\sim 2E_{\rm GR}$ where the cross section is largest, this is different from the smoother limits from IceCube extremely-high-energy (EHE) neutrino flux mainly determined by the shape of the EHE measurement~\cite{Arguelles:2022nbl,IceCube:2018fhm}.}

The constraints on ADM lifetime obtained with the IceCube GR dataset are shown in the bottom two panels of Fig.~\ref{fig:lifetime_limits} for the representative decay modes listed in Table~\ref{tab:decay}. For B1 (therefore also B2) and B3, when the DM mass is around the GR energy, the constraint for the antineutrino decay $L_X<0$ is about two orders of magnitude stronger than $L_X>0$. The difference shrinks as the ADM mass increases and eventually the two branches intersect at $m_X\sim 100~\rm{PeV}$. For $m_X\gtrsim~\rm{EeV}$ the difference between $L_X<0$ and $L_X>0$ decay diminishes, and constraints on ADM become indistinguishable from those on symmetric particles, consistent with the $\bar{\nu}_e$ spectrum in Fig.~\ref{fig:flux_all}. On the other hand, the constraints on $L_X<0$ and $L_X>0$ decay branches in B4 are always very similar regardless of $m_X$, as the decay spectra are less distinguishable. Scenarios with even more symmetric final states should in turn make any $CP$ asymmetry more difficult to identify. {In all these scenarios, the data has a preference on DM decay for $m_X\gtrsim 10$~PeV, although the preference is very mild with $\sim$1.2$\sigma$ significance. Including the IceCube HESE and EHE measurements may improve the limits at low and high masses away from the GR energy, though they are not sensitive to the sign of $L_X$, as detailed in the Supplementary Material.}

At high energies, electroweak corrections also give rise to photons that can be observed at gamma-ray telescopes; these spectra are of course insensitive to the $\nu:\bar \nu$ ratio. In the panels of Fig.~\ref{fig:lifetime_limits} that depict two-body final states, we also show the derived gamma-ray constraints from  LHAASO~\cite{LHAASO:2022yxw} and KASCADE-Grande  \cite{KASCADEGrande:2017vwf,Arguelles:2022nbl}. These gamma-ray constraints are up to an order of magnitude stronger for masses above 100~PeV. However, this channel remains insensitive to the sign of the $B-L$ number and hence to any asymmetry in the dark sector.

The projected sensitivities with 10~yr IceCube-Gen2 exposure are also shown in Fig.~\ref{fig:lifetime_limits} as dashed lines. They have similar mass dependence as the current constraints, achieving a factor of 5 improvement in sensitivity.  We find that the lifetime sensitivity scales with the exposure by the relation $\tau_X\propto \mathcal{E}^{0.59}$ where $\mathcal{E}$ is the exposure. The combination of future neutrino telescopes will of course further improve the sensitivity.

The constraints on decaying dark matter have mild dependence on the neutrino spectral index $\gamma$ and the $\bar{\nu}_e$ fraction $f_{\bar{\nu}_e}$ of the diffuse astrophysical flux. As displayed in Fig.~\ref{fig:flux_all}, current measurements of the diffuse astrophysical flux report spectral indices vary from 2.37 (through-going muons) to 2.87 (HESE)~\cite{IceCube:2020wum,IceCube:2021uhz}. Figure~ \ref{fig:lifetime_limits} assumes $\gamma = 2.87$. Using $\gamma = 2.37$ gives a fivefold increase in the astrophysical flux in the GR energy range, leaving less room for a DM signal and an improvement of $\sim 2$ in the limits. Future projections work in the opposite direction, as expected data would give a larger background, and thus a \textit{lower} sensitivity if the astrophysical spectrum is harder.

In typical astrophysical production scenarios, the fraction of electron antineutrinos $f_{\bar{\nu}_e}$ can range from 0 to $\lesssim 20\%$~\cite{Biehl:2016psj}. We find that varying $f_{\bar{\nu}_e}$ from $1/6$ to 0 leads to a factor  $\sim$1.2 decrease in the lifetime constraints and a factor $\sim$5 improvement in the lifetime sensitivity. 

Finally, we note that if $X$ discussed here is only a fraction $f_X$ of the cosmological DM, then these constraints simply scale to $\tau_X \rightarrow f_X \tau_X$ since the neutrino flux scales with the DM density.

The exact relation between lifetimes constrained here and other observables such as relic abundance depend on which of the operators in Table \ref{tab:decay} are present in the full theory, as well as on other details of the UV completion (see. e.g. \cite{Zhao:2014nsa}). These may also lead to other observables, such as neutrino-dark matter elastic scattering at high \cite{Cline:2022qld,Ferrer:2022kei,Arguelles:2017atb} and low \cite{Escudero:2015yka,Crumrine:2024sdn} energies. The interplay of these constraints with the ones we have presented in this work will be, once again, model-dependent.

\textbf{\textit{Conclusions}} ---
In this work, we have exploited the large increase in neutrino-electron cross section at the Glashow resonance to place leading limits on dark matter decays to neutrino pairs, and have placed the first constraints on heavy ADM decaying to neutrinos. As GR provides unique opportunities for disentangling neutrinos and antineutrinos, current IceCube data sets the ADM lifetime to be longer than $10^{27}-10^{29}$~s for $m_\chi\gtrsim 10$~PeV depending on the decay modes. Thanks to the GR, constraints place here are competitive with---and in some ranges, stronger than---those from gamma ray searches for decay to electromagnetically charged final states such as $\bar b b$, $\tau^+\tau^-$, or $e^+e^-$ \cite{LHAASO:2022yxw}. Even stronger sensitivities will be achieved in the near future with the upcoming generation of neutrino telescopes which provide considerably larger exposure than IceCube.

Distinct from charged cosmic rays, neutrinos directly point back to their sources, and unlike gamma rays, they travel without being attenuated and carry information about the initial $CP$ state. A strong case for complementarity exists: because dark matter in this mass range would produce a flux of gamma rays and neutrinos, a simultaneous measurement of both of these signals would be the only way to pin down the existence and asymmetry of the dark sector, and map out its distribution in the local cosmos.

\begin{acknowledgments}
The authors would like to thank Joseph Bramante and Yue Zhao for helpful discussion. NS is supported by the National Natural Science Foundation of China (NSFC) Project Nos. 12347105, 12475110, 12441504 and 12447101. QL and ACV are supported by the Arthur B. McDonald Canadian Astroparticle Physics Research Institute, with equipment funded by the Canada Foundation for Innovation and the Province of Ontario, and housed at the Queen’s Centre for Advanced Computing. Research at Perimeter Institute is supported by the Government of Canada through the Department of Innovation, Science, and Economic Development, and by the Province of Ontario. ACV is also supported by NSERC, and the province of Ontario via an Early Researcher Award. 

\end{acknowledgments}

\bibliography{ref}

\end{document}


\maketitle
\onecolumngrid
\begin{center}
\textbf{\large Constraints on Heavy Asymmetric and Symmetric Dark Matter  \\
from the Glashow Resonance}

\vspace{0.05in}
{ \it \large Supplementary Material}\\ 
\vspace{0.05in}
{Qinrui Liu, Ningqiang Song and Aaron C. Vincent}
\end{center}
\onecolumngrid
\setcounter{equation}{0}
\setcounter{figure}{0}
\setcounter{section}{0}
\setcounter{table}{0}
\setcounter{page}{1}
\makeatletter
\renewcommand{\theequation}{S\arabic{equation}}
\renewcommand{\thefigure}{S\arabic{figure}}
\renewcommand{\thetable}{S\arabic{table}}
\vspace{-3mm}

\section{Neutrino distribution function in dark matter decay}
\label{app:decaydistribution}

Here we discuss the initial spectra of the neutrino production via our benchmark decay modes. In a two-body decay, the monoenergetic initial states from the decay can be described by a delta function $df_i/dy = \delta(y-y_{i,0})$. Here $y$ represents the energy fraction taken away by particle $i$. For benchmarks B1 and B2, when $m_\psi\ll m_X$, $y_{i,0}=1/2$ for all decay products. If $m_\psi\lesssim m_X$ instead, $y_{\nu,0}=(m_X^2-m^2_{\psi})/(2m_X^2)$. Therefore, the B1 neutrino spectrum matches the spectrum of B2 at mass $m'_X=2y_0m_X$ with the normalization reduced by a factor of 2.

For a multi-body decay mode $X\rightarrow \nu + 2n\psi$, in the massless limit $m_\psi\ll m_X$, the distribution function of $\nu$ is~\cite{Hiroshima:2017hmy}
\begin{align}
    \frac{df_\nu}{dy} = 4N(N-1)(N-2)y^2(1-2y)^{N-3},
\label{eq:nu2nS}
\end{align}
where $N\equiv 3n+1$ and $y\in[0,\frac{1}{2}]$. The distribution function of each product $\psi$ in the pairs is 
\begin{align}
    \frac{df_\psi}{dy}= 8N(N-1)(N-2)(1-2y)^2(4y-1)^{N-3},
\label{eq:nu2nSM}
\end{align}
where $y\in [\frac{1}{4},\frac{1}{2}]$. In benchmark B3, $\psi$ is a dark particle, and it does not contribute to the neutrino flux. In benchmark B4, $\psi=\nu$ instead, and the $\nu$($\bar{\nu}$) distribution is obtained by adding up Eq.~\eqref{eq:nu2nS} and Eq.~\eqref{eq:nu2nSM}. As no matter whether $L_X$ is negative or positive, the neutrino pair contributes the same to the spectrum, making the final spectra close to each which weakens the power of GR. It is also possible that the $n$ $\psi$ pairs are other SM particle pairs, and they can eventually produce neutrinos through hadronization, decay or radiative processes. The exploration of this scenario is left for future work.

\section{Sensitivity of dark matter decay with next-generation neutrino telescopes}
\label{app:projection}

As discussed in the main text, we use the partially-contained event selection for future projections. The expected number of (partially) contained events depends on the the detector effective volume and exposure time of the experiment. Therefore, the partially-contained event rate at a future neutrino telescope with a instrumental volume $V$ can be scaled from that of IceCube ($\sim 1~\rm{km^3}$ in size) by a factor of $\sim 2\left(V/1~\rm{km^3}\right)$. We use the public Monte Carlo (MC) simulation of HESE~\cite{IceCube:2020wum}\footnote{\href{https://github.com/icecube/HESE-7-year-data-release}{https://github.com/icecube/HESE-7-year-data-release}} to compute the latter, which includes reconstruction effects such as energy reconstruction and misidentification, as well as ice and detector systematics. Since the HESE MC does not account for Doppler broadening and initial photon radiation, we include these with an appropriate reweighting.

The Glashow resonance gives rise to a distinct signal compared with DIS: mesons decay into muons with energies of tens of GeV whose Cherenkov photons are visible as early pulses. This distinguishes them from a charged current DIS cascade~\cite{IceCube:2021rpz}.  Neutral current DIS events are also expected to create hadronic cascades, but they require the incoming neutrino to be much more energetic to deposit similar amount of energy in the detector. This is suppressed due to the low astrophysical flux expected at higher energies. The $W$ decay channel to muons  results in a track without a cascade at the primary vertex, in contrast to charged current $\nu_\mu$ DIS interactions in which the nucleon is disintegrated. This contribution is nevertheless subdominant as the branching ratio to muons rate is much lower than to hadrons.

In order to compute the sensitivity, we perform a binned likelihood ratio analysis based on the Poisson likelihood of Glashow-induced cascades (cas) and tracks (tr) in each of the three bins defined in Ref.~\cite{IceCube:2021rpz}:
\begin{equation}
\mathcal{L} = \prod_i^3 \mathcal{L}^{\mathrm{cas}}_i\prod_j^3 \mathcal{L}^{\mathrm{tr}}_j =\prod_i^3\frac{\mu^\mathrm{cas}_i}{k^{\mathrm{cas}}_i!}e^{-\mu^{\mathrm{cas}}_i}\prod_j^3\frac{\mu^{\mathrm{tr}}_j}{k^{\mathrm{tr}}_j!}e^{-\mu^{\mathrm{tr}}_j}.
\label{eq:likelihood}
\end{equation}
Here, $\mu$ reprensents expected events per bin, and $k$ correspond to the observed GR counts. In each case, $\mu_{i} = \mu^{\mathrm{DM}}_{i} + \mu^{\mathrm{astro}}_{i}$ consists of contributions from the DM and the diffuse astrophysical neutrino flux and can be computed as
\begin{equation}
 \mu^{k}_i  =   T\int_{\Delta E_i} \int_{E_\nu} A_{\mathrm{eff}}(E_\nu)\frac{d\Phi^{k}_\nu}{dE_\nu}\frac{d\mathcal{P}}{dE_{\mathrm{reco}}}dE_\nu dE_{\mathrm{reco}}\,,
\label{eq:event_number}
\end{equation}
where $k=\mathrm{DM}$ or $\mathrm{astro}$, and $T$ is the exposure time of the experiment. The effective area $A_{\rm eff}$ depends on the interaction type, the neutrino energy and the zenith angle. Since we are considering a direction-averaged flux, $A_{\mathrm{eff}}$ here is also averaged in zenith angle. $d\mathcal{P}/dE_{\mathrm{reco}}$ takes into account the reconstruction effect, indicating the probability of observing an event of $E_{\mathrm{reco}}$ from a incoming neutrino with energy $E_\nu$. $\Delta E_i$ is the width of the reconstructed energy in bin $i$. 

\begin{figure}[t!]
    \centering
\includegraphics[width=0.5\columnwidth]{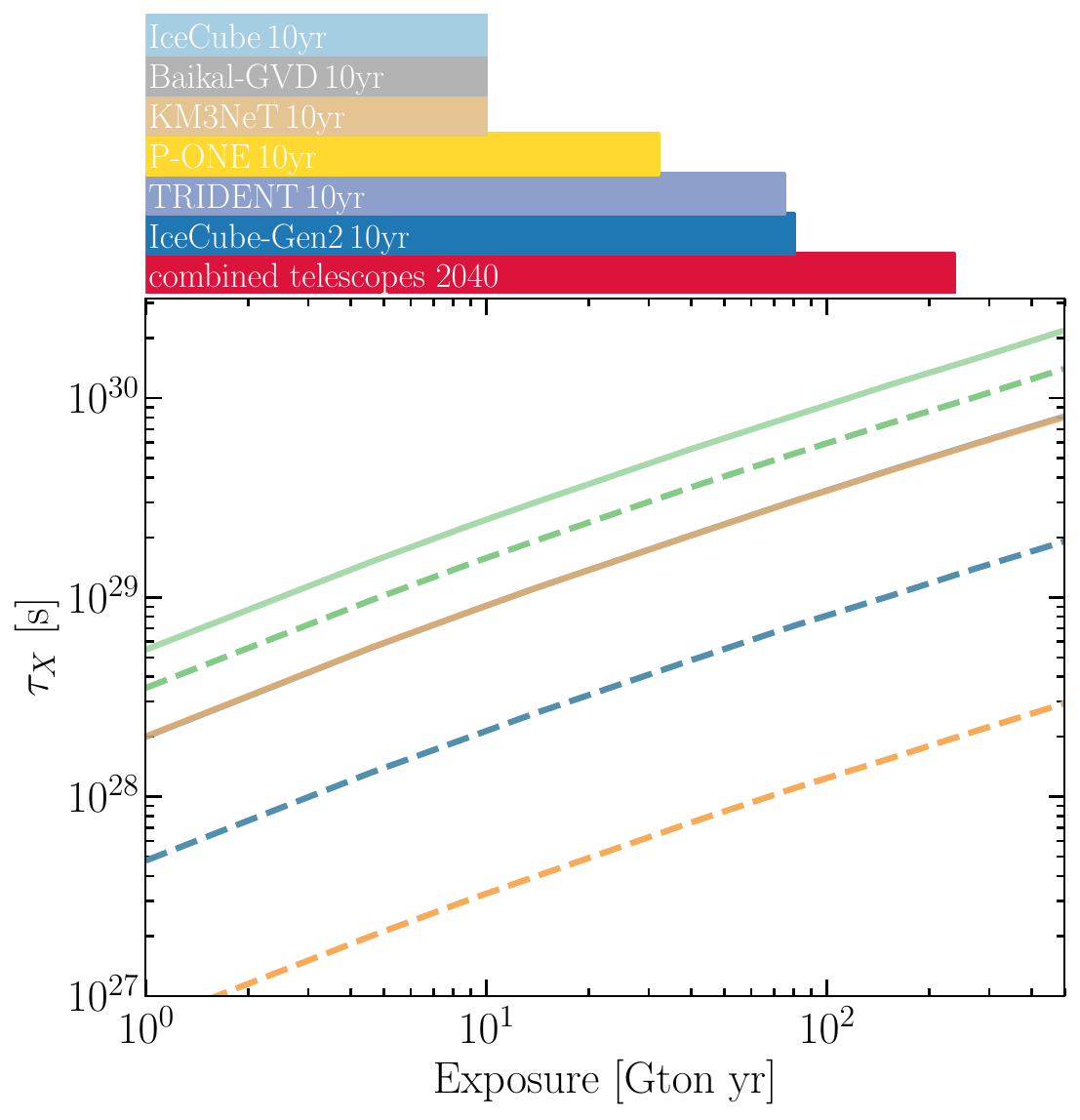}
    \centering
    \vspace*{-0.5cm}
    \hspace*{0.9cm}
\includegraphics[width=0.5\columnwidth]{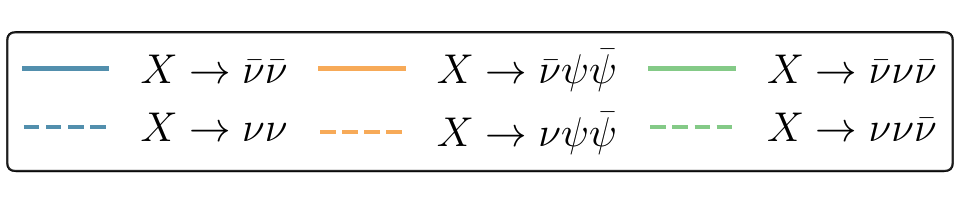}
    \caption{The sensitivity evolution with the exposure for the three different decay modes for $m_X=20~\rm{PeV}$. Solid lines correspond to negative lepton numbers and dashed lines correspond to positive lepton numbers (the $X\rightarrow \bar{\nu}\bar{\nu}$ line accidentally coincides with $X\rightarrow \bar{\nu}\psi\bar{\psi}$ at $m_X=20~\rm{PeV}$). The bars on the top indicate the 10~yr exposure of IceCube and next-generation neutrino telescopes. The red bar corresponds to an estimated exposure in year 2040 with combined neutrino telescopes which are composed of 19~yr IceCube, 15~yr Baikal-GVD+KM3NeT and 10~yr P-ONE+TRIDENT+IceCube-Gen2.}
    \label{fig:exposure_limits}
\end{figure}

Fig.~\ref{fig:exposure_limits} shows the evolution of the sensitivity for $m_X=20$~PeV as the exposure grows for the representative decay channels. The projected sensitivity approximately follows $\tau_X/\tau_0\simeq \mathcal{E}^{0.59}$, where $\mathcal{E}$ is the exposure of the experiment and $\tau_0$ depends on the channel. For B2 $X \rightarrow \bar{\nu}\bar{\nu} (\nu \nu)$, $\tau_0\simeq 10^{28.3} (10^{27.7})$~s. For B3 $X \rightarrow \bar{\nu}(\nu)+\psi\bar{\psi}$, $\tau_0\simeq 10^{28.3} (10^{26.9})$~s. For B4 $X \rightarrow \bar{\nu}(\nu)+\nu\bar{\nu}$, $\tau_0\simeq 10^{28.8} (10^{28.6})$~s.

At the top, the approximate 10~yr exposure timelines of next-generation telescopes are marked. These are mainly based on sensitivities presented before full experimental details or funding were finalized. We keep these numbers here, as up-to-date sensitivities have not been published. In addition to IceCube-Gen2, the high-energy module of KM3NeT~\cite{KM3Net:2016zxf} will be instrumented with two 115-string arrays, which in combination provides an effective volume comparable to IceCube. Although Baikal-GVD has been taking data since 2018~\cite{Baikal-GVD:2018isr}, the detector is expected finally reach an instrumented volume of $\sim$1~$\mathrm{km^3}$. P-ONE~\cite{P-ONE:2020ljt}, a planned water-Cherenkov detector in the north-eastern Pacific Ocean, will be constructed with an estimated volume of 3.2~$\mathrm{km^3}$, although more recent conference presentations suggest a smaller volume  \cite{Twagirayezu:2023cpv}. TRIDENT~\cite{Ye:2022vbk}, proposed to be built in the South China Sea, is partially funded and has an ultimate goal of reaching 7.5~$\mathrm{km^3}$. 
For the combined exposure of neutrino telescopes by 2040, it is based on the approximate timeline that the full configuration of Baikal-GVD and KM3NeT will start taking data in 2025 while IceCube-Gen2, P-ONE and TRIDENT will be turned on in 2030, corresponding to an exposure of 19~yr for IceCube, 15~yr for Baikal-GVD and KM3NeT, and 10~yr for P-ONE, TRIDENT and IceCube-Gen2 respectively. The volumes and timelines we have adopted for upcoming detectors are tentative and subject to changes, and may be augmented by the instrumentation of even larger detectors, such as the recently proposed HUNT~\cite{Huang:2023mzt}. As the normalization of the lifetime sensitivity scales with the exposure while the shapes approximately keep the same with varying mass, the projected sensitivity lines in the main text can be scaled to obtain the sensitivity lines with any more realistic exposure in the future. 

\section{Effects of corrections on the GR cross section}\label{app:xsec}

Following the increasing number of studies on high-energy neutrino interactions, subleading effects are being considered when computing cross sections. For GR cross sections, the discussed corrections are from the Doppler broadening~\cite{Loewy:2014zva} and the initial photon radiation~\cite{Gauld:2019pgt,Garcia:2020jwr}. These effects lead to a $\sim$30\% drop at the resonant energy, a slight broadening and higher values above the resonant energy of the cross section~\cite{Huang:2023yqz}. The former has been implemented in the IceCube GR analysis~\cite{IceCube:2021rpz}. As the effective area of contained or partially contained events is approximately proportional to the cross section, the effect on the event rate can be estimated. Here, with the HESE MC, we obtain the effective area incorporating the two effects by scaling the MC weight of each simulated event by the ratio of the modified cross section and the original cross section. With the 3 energy bins discussed above, the first 2 bins have a negligible change with a drop by
less than several percent while for the bin above the resonant energy there is a 20\%-30\% increase for a power-law spectrum.

We check the effect of the cross section corrections on the constraints on ADM, and find an improvement at only a level of several percent in the DM lifetime, which is negligible. The improvement is less significant compared to the change in the event rate, since the GR event rates from DM decay and diffuse astrophysical neutrino flux both change similarly due to the cross section corrections.

\section{Limits on the lifetime of dark matter with high-energy starting events and extremely-high-energy neutrino flux measurements}

\begin{figure}[t!]
    \subfigure
    {\includegraphics[width=0.48 \linewidth]{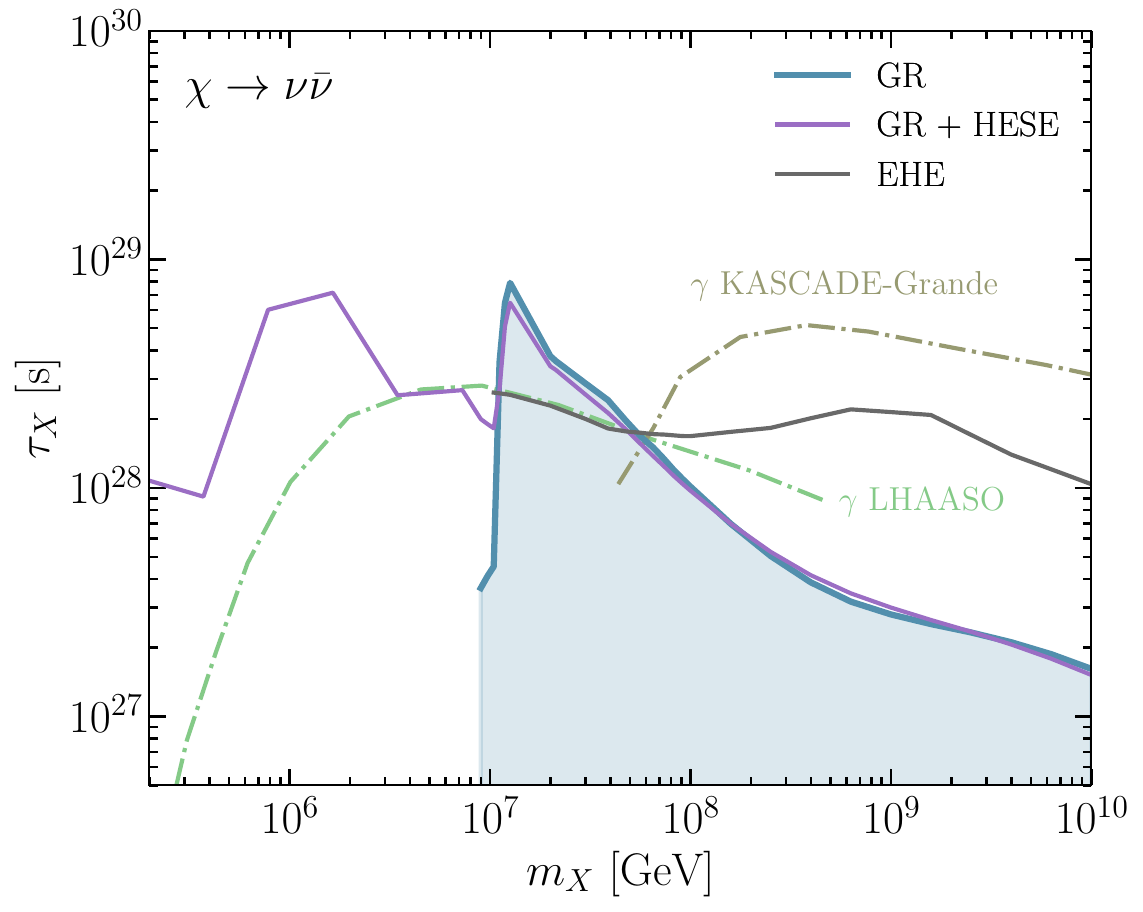}}
    \subfigure
    {\includegraphics[width=0.48\linewidth]{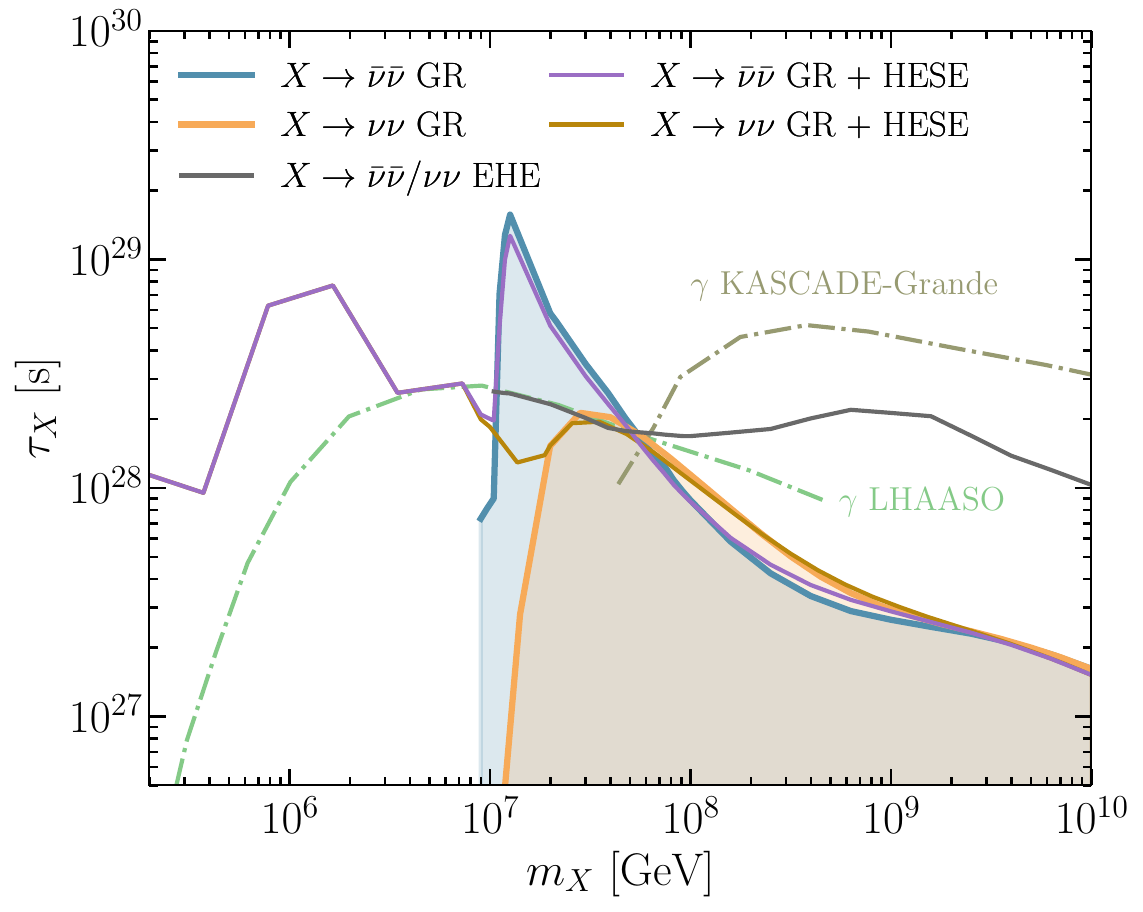}}\\
    \subfigure
    {\includegraphics[width=0.48\linewidth]{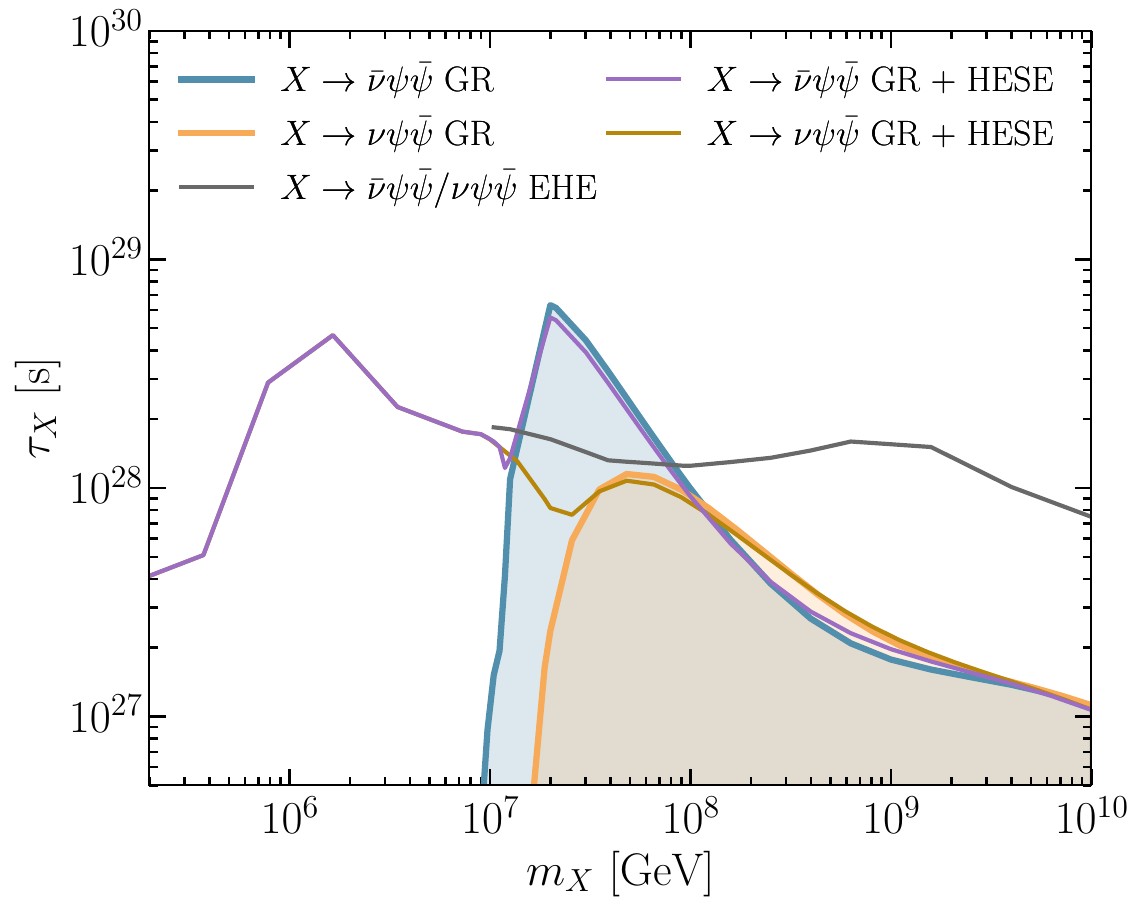}} 
    \subfigure
    {\includegraphics[width=0.48\linewidth]{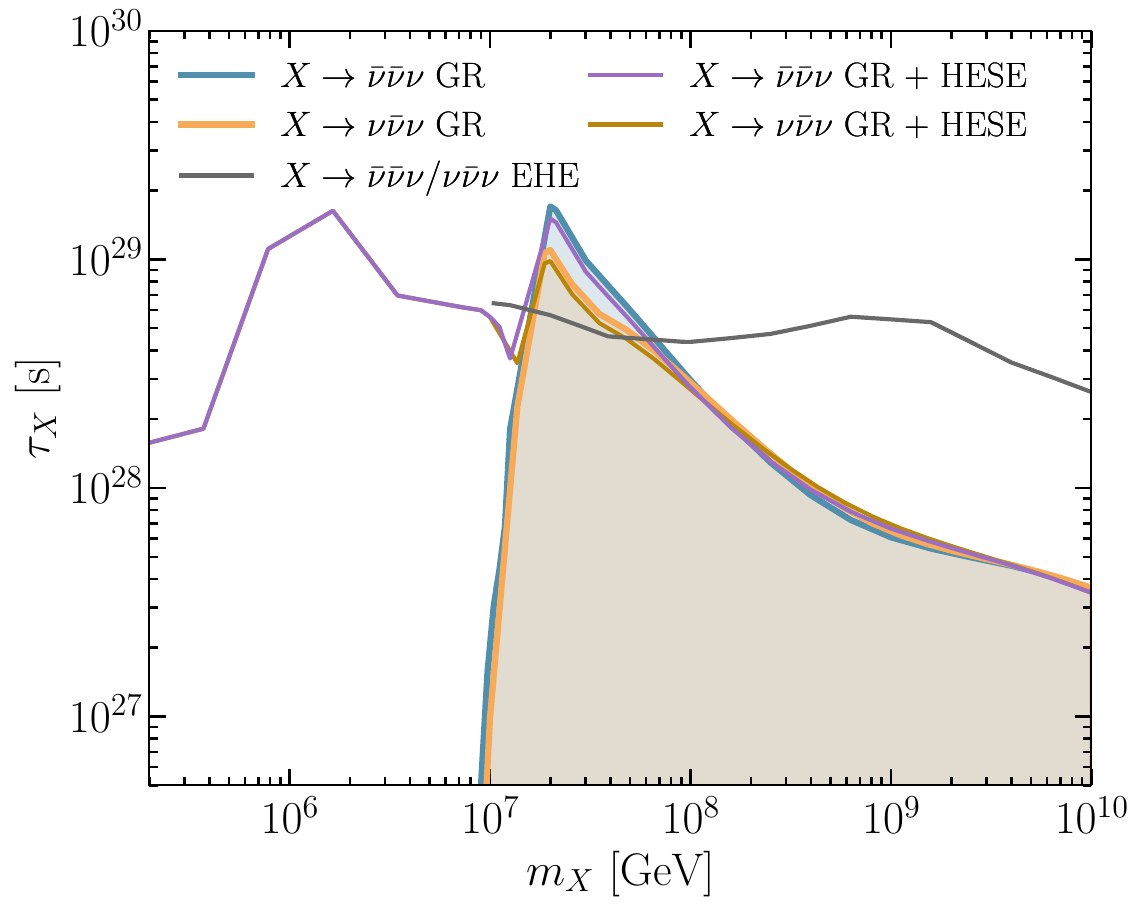}}  
    \caption{Comparison between the 90\% C.L. constraints (blue and orange shaded regions) shown in Fig.~2 in the main text and combined constraints including current HESE and GR observations (solid purple and brown thin lines) and constraints with the EHE upper limit (solid gray thin lines). HESE and EHE provide leading constraints for $m_X\lesssim 10$~PeV and $m_X\gtrsim 100$~PeV respectively, while the sign of the lepton number is degenerated as $\nu$ and $\bar{\nu}$ are not differentiated without the contribution from GR. GR stands out when the ADM mass is in between 10~PeV$\sim$100~PeV.}
    \label{fig:compare_lifetime_limits}
\end{figure}

We also show the limits on the lifetime of DM with HESE and EHE neutrino flux measurements~\cite{IceCube:2020wum,IceCube:2018fhm} in order to present limits for a wider mass range as well as to provide a comparison to the limits obtained with merely GR observation. To be noticed that HESE and EHE include neutrino events of all flavors and it is difficult to differentiate neutrinos and antineutrinos at high energies outside the GR energy window. Here, we combine the HESE measurements, \ie~the frequentist analysis results reported in Table G.1. in Ref.~\cite{IceCube:2020wum}, with GR observation by adding the segmented power-law fluxes into the likelihood function Eq.~4 used in this work. Since the HESE flux is the all-flavor neutrino flux, the DM flux and astrophysical diffuse flux are also replaced by the all-flavor neutrino flux in the likelihood for the HESE part. 
We remove the last segment with no event observation, the energy bin of which overlaps partially with the GR observation.
Since the energy range of HESE is down to 60~TeV, the DM mass we consider starts from $\sim100$~TeV. The combined results are shown in Fig.~\ref{fig:compare_lifetime_limits} as thin purple and brown lines. It can be seen that for ADM scenarios, for the mass range presented in this work, \ie~$m_X\gtrsim$10~PeV, combining HESE results in consistent constraints with the results computed with only the GR observation except for $X\rightarrow \nu\nu$ and $X\rightarrow \nu\psi\bar{\psi}$ where HESE improves the constraints at the lower mass end. This is because the $\bar{\nu}$ spectra for these two modes peak at energies slightly lower than $m_X/2$, as can be seen in Fig.~1, where HESE complements the loss of sensitivity for GR towards this energy range. 
For lower DM mass, GR no longer contributes to the measurement, and hence the constraints for both lepton number signs converge.

Regarding the fitting significance, we find including DM results in the test statistics TS$\sim$1.4 when using GR only, corresponding to $\sim 1.2 \sigma$. After combining HESE, the TS value is comparable when $m_X \sim 10~\rm{PeV}-20~\rm{PeV}$ and becomes slightly smaller for heavier DM.

The EHE measurement reports the differential 90\% C.L. flux upper limit above 5~PeV~\cite{IceCube:2018fhm}. Here, we use the upper limit to constrain the total neutrino flux independently from HESE and GR observations. We compute the constraints by straightforwardly comparing the integrated flux in a specific bin. In order to provide the most conservative constraints, we do not include any background flux from Standard Model astrophysical sources. The EHE differential flux limit is set assuming $E_\nu^{-1}$ spectrum with an interval of one decade, \ie~$d\Phi_{\rm{lim}}/dE_\nu \propto E_\nu^{-1}$. Therefore, for each $m_X$, we compute the $\tau_{X}$ satisfies
\begin{align}
\begin{split}
    \Phi_{\rm{lim}} &= \int_{E_\nu\cdot 10^{-\frac{1}{2}}}^{E_\nu\cdot 10^{\frac{1}{2}}} \frac{d\Phi_{\rm{lim}}}{dE_\nu}\left(E_\nu\right)dE_\nu = \frac{d\Phi_{\rm{lim}}}{dE_\nu}E_\nu \rm{ln}10 \\
    &=\int_{E_\nu\cdot 10^{-\frac{1}{2}}}^{E_\nu\cdot 10^{\frac{1}{2}}} \frac{d\Phi^{\rm{DM}}_{\nu}}{dE_\nu}\left(E_\nu;\tau_{X}\right)dE_\nu. 
\end{split}
\end{align}
The results are shown as gray lines in Fig.~\ref{fig:compare_lifetime_limits}. For the two-body decay modes, our limits are comparable to the EHE limits reported in Ref.~\cite{Arguelles:2022nbl} for $X\rightarrow \nu\bar{\nu}$. It can be seen that EHE offers improved constraints when $m_X\gtrsim 100$~PeV. However, similar to HESE, $\nu$ and $\bar{\nu}$ cannot be differentiated to obtain different limits for 
different lepton number signs. 

Overall, we find that GR is advantageous when ADM mass is around 10~PeV$\sim$100~PeV, in particular, GR always leads to the most stringent limits when the lepton number is negative.

\bibliography{ref}